\documentclass[sigconf,natbib]{acmart}

\AtBeginDocument{%
  }

\copyrightyear{2023} 
\acmYear{2023} 
\setcopyright{rightsretained} 
\acmConference[ICTIR '23]{Proceedings of the 2023 ACM SIGIR International Conference on the Theory of Information Retrieval}{July 23, 2023}{Taipei, Taiwan}
\acmBooktitle{Proceedings of the 2023 ACM SIGIR International Conference on the Theory of Information Retrieval (ICTIR '23), July 23, 2023, Taipei, Taiwan}
\acmDOI{10.1145/3578337.3605135}
\acmISBN{979-8-4007-0073-6/23/07}

\makeatletter
\gdef\@copyrightpermission{
 \begin{minipage}{0.3\columnwidth}
  \href{https://creativecommons.org/licenses/by/4.0/}{\includegraphics[width=0.90\textwidth]{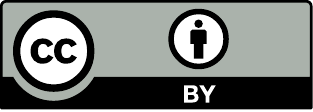}}
 \end{minipage}\hfill
 \begin{minipage}{0.7\columnwidth}
  \href{https://creativecommons.org/licenses/by/4.0/}{This work is licensed under a Creative Commons Attribution International 4.0 License.}
 \end{minipage}
 \vspace{5pt}
}
\makeatother

\usepackage{subcaption}
\usepackage{graphicx}
\usepackage{url}
\usepackage{booktabs}
\usepackage{multirow}
\usepackage{enumitem}

\begin{document}

\title{Outcome-based Evaluation of Systematic Review Automation}

\author{Wojciech Kusa}
\email{wojciech.kusa@tuwien.ac.at}
\orcid{0000-0003-4420-4147}
\affiliation{%
  \institution{TU Wien}
  \city{Vienna}
  \country{Austria}
}

\author{Guido Zuccon}
\email{g.zuccon@uq.edu.au}
\orcid{0000-0003-0271-5563}
\affiliation{%
  \institution{The University of Queensland}
  \city{Brisbane}
  \country{Australia}
}

\author{Petr Knoth}
\email{petr.knoth@open.ac.uk}
\orcid{0000-0003-1161-7359}
\affiliation{%
  \institution{The Open University}
  \city{Milton Keynes}
  \country{The United Kingdom}
}

\author{Allan Hanbury}
\email{allan.hanbury@tuwien.ac.at}
\orcid{0000-0002-7149-5843}
\affiliation{%
  \institution{TU Wien}
  \city{Vienna}
  \country{Austria}
}

\begin{abstract}
Current methods of evaluating search strategies and automated citation screening for systematic literature reviews typically rely on counting the number of relevant publications (i.e. those to be included in the review) and not relevant publications (i.e. those to be excluded). Significant importance is put into promoting the retrieval of all relevant publications through great attention to recall-oriented measures, and demoting the retrieval of non-relevant publications through precision-oriented or cost metrics. This established practice, however, does not accurately reflect the reality of conducting a systematic review, because not all included publications have the same influence on the final outcome of the systematic review. More specifically, if an important publication gets excluded or included, this might significantly change the overall review outcome, while not including or excluding less influential studies may only have a limited impact. However, in terms of evaluation measures, all inclusion and exclusion decisions are treated equally and, therefore, failing to retrieve publications with little to no impact on the review outcome leads to the same decrease in recall as failing to retrieve crucial publications.

We propose a new evaluation framework that takes into account the impact of the reported study on the overall systematic review outcome. We demonstrate the framework by extracting review meta-analysis data and estimating outcome effects using predictions from ranking runs on systematic reviews of interventions from CLEF TAR 2019 shared task. We further measure how closely the obtained outcomes are to the outcomes of the original review if the arbitrary rankings were used. We evaluate 74 runs using the proposed framework and compare the results with those obtained using standard IR measures. We find that accounting for the difference in review outcomes leads to a different assessment of the quality of a system than if traditional evaluation measures were used. Our analysis provides new insights into the evaluation of retrieval results in the context of systematic review automation, emphasising the importance of assessing the usefulness of each document beyond binary relevance.

\end{abstract}



\maketitle

\section{Introduction}
A systematic literature review is a well-established and rigorous methodology for synthesising and evaluating the evidence on a specific research question, which is particularly important in the field of medicine~\cite{Jo2009}. 
However, it is also gaining importance in other areas such as social sciences and engineering~\cite{castillo2022apisser,petticrew2008systematic,keele2007guidelines,bilotta2014use}.
The process involves a systematic search, critical appraisal, and synthesis of the available literature on a topic. During the critical appraisal step, every included publication has its weight and effect calculated based on the outcomes reported by that publication. This information influences the final outcome of the review.

One of the essential steps in conducting a systematic review is the process of citation screening, in which a large number of publications are initially identified through a literature search and then screened to determine those relevant to the review \cite{Bannach-Brown2019,Tsafnat2018AutomatedCharacteristics}. This process can be time-consuming and labour-intensive, involving making thousands of eligibility decisions. Given the importance of citation screening in systematic literature reviews, there have been numerous attempts to automate the process~\cite{o2015using}. Previous studies have investigated the use of automated citation screening methods for systematic literature reviews by utilising various natural language processing (NLP), machine learning (ML), and information retrieval (IR) methods to retrieve, rank, or classify references~\cite{o2015using,VanDinter2021,Kusa2022AutomationStudy,Howard2016a,scells2020automatic,Kanoulas2019CLEF2T,scells2020you,scells2021comparison,wang2022automated,wang2023neural,wang2023can}. 

To understand the effectiveness of automated citation screening methods, practitioners have relied on metrics based on the notions of recall, precision and cost -- and of a binary assessment of relevance\footnote{Every publication to be included in the review is labelled as relevant, while every excluded publication is non-relevant.}~\cite{Kusa2022AutomationStudy,VanDinter2021,o2015using}. This practice assigns to every publication to be included in the review the same importance. 
So, for example, if method $M_1$ identifies as potentially relevant publications $\{A, B, C\}$ while method $M_2$ identifies publications $\{A, D, E\}$, and the ground truth is that the relevant publications are $\{A, B, D\}$, then $M_1$ and $M_2$ achieve the same recall, precision and cost. 
However, we argue, that the two sets $\{A, B, C\}$ and $\{A, D, E\}$ may not be equally important, and thus identifying either of $B$ or $D$ may not be equivalent if the outcomes of the review were considered. 
In fact, if excluded, some publications can significantly change a review's conclusion to the extent that the conclusion might be the opposite (e.g., from favouring a drug to favouring a placebo)~\cite{NUSSBAUMERSTREIT20181,nussbaumer2020excluding}. 
On the other hand, not including other publications might have only a small quantitative impact on the outcomes of the review. 

We argue that a holistic evaluation of retrieval and automated citation screening methods for systematic review creation should not only consider the concepts of recall, precision and cost, but also the quality of the outcomes generated from the analysis of the automatically included publications.
Following this direction, we propose a new evaluation framework that considers inclusion and exclusion information and meta-analysis data from reviews created by Cochrane -- the largest organisation responsible for creating systematic literature reviews in medicine,\footnote{\url{https://www.cochrane.org}} to estimate outcomes and weights of included publications.
This information can be used to assess the quality of ranking and classification methods. 
This framework allows for assessing automatic approaches from the angle of how closely their \emph{outcomes} -- not just their set of included publications -- are to the outcomes of the original review. 
By comparing the outcomes of the automated model to those of the original review, we can gain a better understanding of the quality of the automated approach and its effect on the final outcome of the review. 

We propose five aspects of analysis focusing on different features of review outcomes.
We explore initial experiments on the CLEF TAR 2019 dataset~\cite{Kanoulas2019CLEF2T}.
Our simulation results show that by randomly removing one publication per review (average recall of 92\% publications), 95\% of outcomes remain unchanged. 
However, after removing five publications (average recall of 63\%), 76\% of the outcomes are still the same, showing that the relationship between recall and achieved outcomes is not linear. 
We also show that the outcome-based evaluation emphasises different aspects of the models' performance than the traditional IR evaluation measures.
We finally propose multi-objective optimisation to handle the problem of non-estimable outcomes.

We believe this new evaluation approach will provide a better understanding of the impact of automatic literature screening methods on the outcome of systematic literature reviews and help identify areas in which these methods can be improved.

\section{Related Work}

The effectiveness of automatic approaches for search strategy creation and systematic review screening has been traditionally evaluated using binary relevance ratings~\cite{o2015using,Kanoulas2019CLEF2T,VanDinter2021}, often sourced at the title and abstract screening level, rather than at the full-text level. %

When the screening problem is treated as a ranking task (e.g., for the sub-task of screening prioritisation or stopping prediction), then rank-based metrics and metrics at a fixed cut-off are commonly used, e.g., $nDCG@n$, $Precision@n$, $Recall@n$, last relevant found~\cite{Scells2017,Howard2020}.
Cost-based and economic-based metrics are also used, especially in the context of the query formulation task in the CLEF TAR shared task \cite{Kanoulas2019CLEF2T,Kanoulas2017CLEFOverview,Kanoulas2018CLEFOverview}, e.g., total cost (TC) or total cost with a weighted penalty (TCW).
The TREC Total Recall track~\cite{grossman2016trec} also used a cut-off based metric, $recall @ aR + b$, which is defined as the recall achieved when $aR + b$ documents have been identified, where $R$ is the number of relevant documents in the collection and $a$ and $b$ are parameters. When $a = 1$ and $ b = 0$, $recall @aR + b$  is equivalent to R-precision. 
In the patent domain, the PRES score has been proposed which takes into account achieved recall and the user’s search effort~\cite{magdy2010pres}.

When the screening problem is treated as a classification task, metrics based on the confusion matrix and the notion of Precision and Recall are commonly used~\cite{VanDinter2021,o2015using}: aside from Precision and Recall, metrics include variations of the harmonised mean between the two, i.e. F$_\beta$--score, utility, U19~\cite{wallace2010semi,Wallace2010ActiveScreening,wallace2011should}, sensitivity-maximising thresholds~\cite{dalal2013pilot}, and AUC~\cite{cohen2010prospective}. 
Another metric, Work Saved over Sampling (WSS), measures the amount of work saved when using machine learning models to screen irrelevant publications \cite{Cohen2006,Matwin2010AReviews,Kontonatsios2020,Kusa2022AutomationStudy}.
The True Negative Rate (TNR) and nreTNR (normalised rectified TNR) were proposed as an alternative as it addresses some of the limitations of WSS regarding averaging scores from multiple datasets \cite{KUSA2023200193,kusa2023vombat}.

\citet{NUSSBAUMERSTREIT20181} compared repeated literature searches using 14 abbreviated approaches (combinations of various databases with and without searches of reference lists) on a sample of 60 Cochrane systematic reviews of clinical interventions.
They re-calculated the main summary-of-findings table of each Cochrane review and asked original review authors whether the conclusions changed compared to the original review.
They found that in only 2\% of cases (95\% CI: 0\%--9\%), combining one database with another or with searches of reference lists was falsely reaching an opposite conclusion compared to comprehensive searches.
This outcome shows that identifying \emph{all} relevant studies is not always crucial for obtaining the same review results.

Automated citation screening has become increasingly popular in systematic literature reviews due to its potential to reduce the time and cost required. 
However, current evaluation methods for these methods are limited to binary relevance assessment, where each publication is considered either relevant or irrelevant, and do not account for the impact of each publication on the review outcome. 
This is a vital issue, as the assumption that all relevant publications are equally important to the final outcome of the systematic review is not necessarily valid. 
Without an accurate assessment of the importance of each document, the conclusions of a systematic review may be biased or incomplete. 
To address this issue, in this paper, we propose a novel methodology for assessing citation screening based on evaluating outcome differences, which enables us to determine the impact of each publication on the systematic review.

\section{Evaluation Framework} \label{sec:evaluation_framework}

This paper proposes a new evaluation framework for automated citation screening. 
Our framework includes three steps which are detailed in the following subsections. 
The first step is data extraction, where we extract statistics of the studies included in the review and match studies to publications. 
The second step is model evaluation, where we use the extracted data to estimate the review's outcomes for rankings or classifications of the citation list. 
The third step is the analysis of the results, where we compare the outcomes obtained from the alternative rankings to the outcomes of the original review.
Our proposed framework allows for a more nuanced evaluation of automated citation screening methods. 
By considering the impact of each publication on the review's outcomes, we can identify which publications are most important to retrieve and prioritise them accordingly.
Next we describe each step in detail.

\subsection{Data Extraction} \label{sec:data_extraction}

Cochrane systematic reviews distinguish between \emph{study} and \emph{publication}. 
A study is a distinct piece of research conducted to answer a specific research question or investigate a particular hypothesis. 
It typically involves a group of participants, data collection methods, and specific objectives.
Publications, on the other hand, are the atomic units which reviewers screen. 
Each study can be reported by several publications, such as journal articles, conference proceedings, or research reports. 
Each publication may present different aspects or findings of the same study, but they are all derived from the same underlying research.
We assume that a study has been found if at least one publication reporting it was successfully retrieved.

For every review, based on its Cochrane review ID, we identify its corresponding RevMan file and list of included publications. 
A RevMan file is the format used by Cochrane containing all statistical data about studies and outcomes included in the review.
Outcomes of Cochrane reviews are reported in the following hierarchy: one comparison can have several outcomes, and one outcome can consist of a few subgroups. 
We extract all metadata from the RevMan files, such as the comparisons, outcomes and subgroups and the results of every included study. 
Note that the use of RevMan files is for experimental convenience, but is not a requirement from the framework: the required data could be provided in other formats.
Furthermore, Cochrane recently announced that future systematic literature reviews would contain statistical data in more common \textsc{csv} and \textsc{ris} formats.\footnote{\url{https://www.cochrane.org/news/cochrane-improving-way-we-manage-and-share-data-associated-our-reviews}}

Cochrane reports a list of included publications and studies which correspond to them.
Traditionally, retrieval was conducted at the level of publications~\cite{Kanoulas2017CLEFOverview,Kanoulas2018CLEFOverview,Kanoulas2019CLEF2T}.
In order to be able to re-use previous relevance judgments, we need to assign PubMed IDs to these publications.
Our process for matching PubMed IDs to publications is based on four steps in the following order:

\begin{itemize}[leftmargin=14pt]
    \item We check if the PubMed ID information is provided on the Cochrane references webpage.
    \item We conduct search in PubMed using \textsc{Entrez}\footnote{\url{https://www.ncbi.nlm.nih.gov/search/}} by searching for the same title and authors.
    \item We search for the PubMed ID in SemanticScholar\footnote{\url{https://www.semanticscholar.org}} using publication DOI from Cochrane references webpage.
    \item We search again in PubMed, this time with a relaxed requirement by searching for an exact match in the title only.
\end{itemize}

\subsection{Model Evaluation}

When conducting a meta-analysis, for every outcome, each study has its weight and effect size calculated first (respectively columns 6 and 7 on example forest plots in Figure~\ref{fig:variants}).
Effect size is an essential statistical concept in the analysis of research data~\cite{higgins2019choosing}. 
It is a measure that quantifies the magnitude of difference between two groups in a study. 
Researchers use a variety of effect measures to compare outcome data between two intervention groups, including odds ratios and mean differences.

For instance, in ratio effect measures, a value of 1 represents no difference between the groups \cite{deeks2019analysing,Deeks2010StatisticalAI}. 
On the other hand, in difference measures, a value of 0 represents no difference between the groups. 
Values higher or lower than these ``null'' values may indicate either benefit or harm of an experimental intervention, depending on the order of the interventions in the comparison and the nature of the outcome.
Every estimate is expressed with a measure of uncertainty, such as a confidence interval (CI) or standard error (SE).

Effects depend on the number of events reported by that study, whereas weights assigned to each study are influenced by other studies included in this outcome.
So when removing one study from the meta-analysis, only the weights of the remaining studies will change, but their effect sizes will stay the same (compare Figures~\ref{fig:res_a} and \ref{fig:res_c}).
There are several types of outcomes reported by Cochrane, in our study, we focus on the dichotomous and continuous outcomes only and calculate them following the approach by \citet{Deeks2010StatisticalAI}.

Our framework takes arbitrary ranking or classification runs and calculates the final outcomes of the review based on publications included in the run. 
When evaluating a classification run or a search result, we take all publications predicted as relevant.
When evaluating ranking runs, we need to assume a cut-off point.
Previous studies working on systematic review automation used either cut-off at r\% of recall~\cite{Cohen2006,KUSA2023200193}, or at d\% of total dataset size~\cite{Kanoulas2017CLEFOverview,Kanoulas2018CLEFOverview}.

\subsection{Results Analysis} \label{sec:results-analysis}

\begin{figure*}[ht]
\centering
\begin{subfigure}[t]{0.48\textwidth}
    \includegraphics[width=\textwidth]{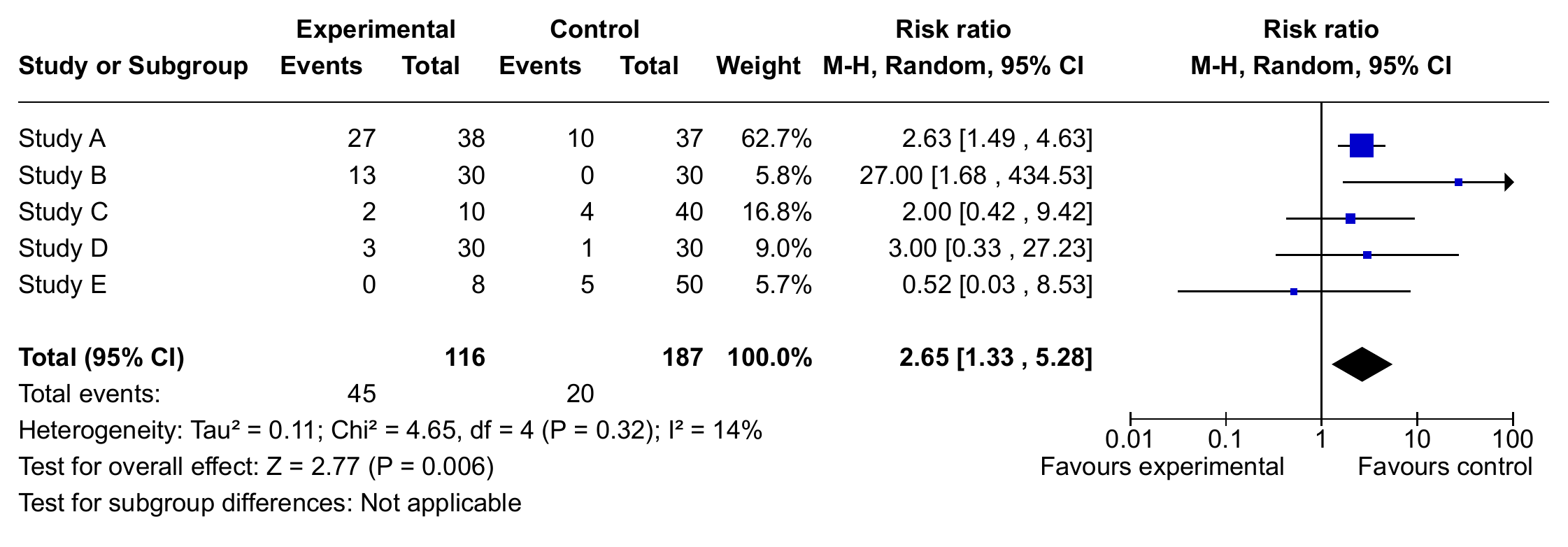}
    \caption{Hypothetical review outcome with 5 included studies.}
    \label{fig:res_a}
\end{subfigure}
\begin{subfigure}[t]{0.48\textwidth}
    \includegraphics[width=\textwidth]{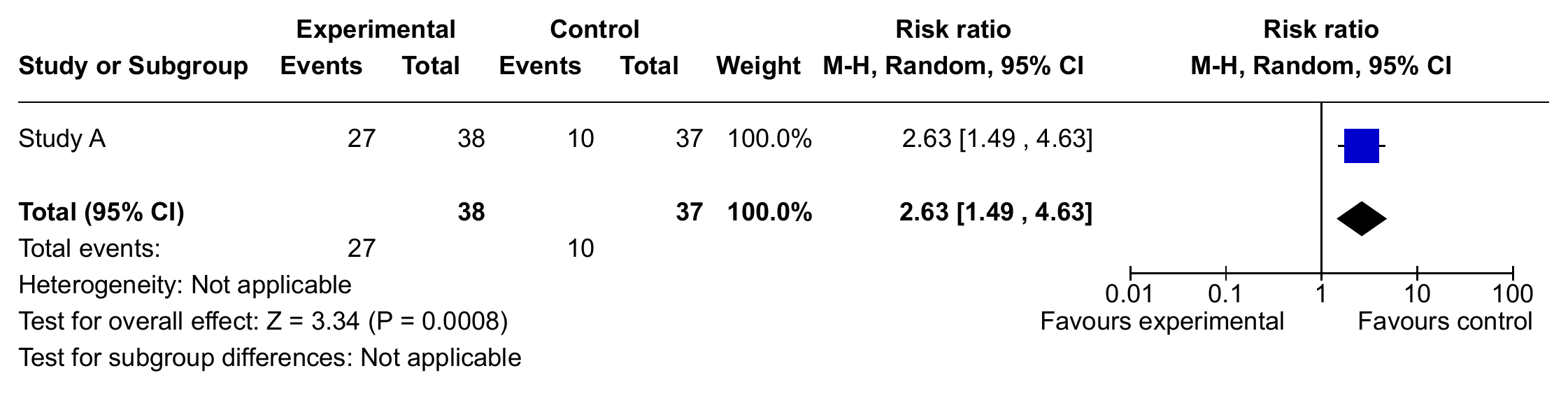}
    \caption{Not including studies B, C, D and E still keep the review outcome approximately the same (absolute difference: 0.02, relative difference: 0.0076).}
    \label{fig:res_b}
\end{subfigure}
\begin{subfigure}[t]{0.48\textwidth}
    \includegraphics[width=\textwidth]{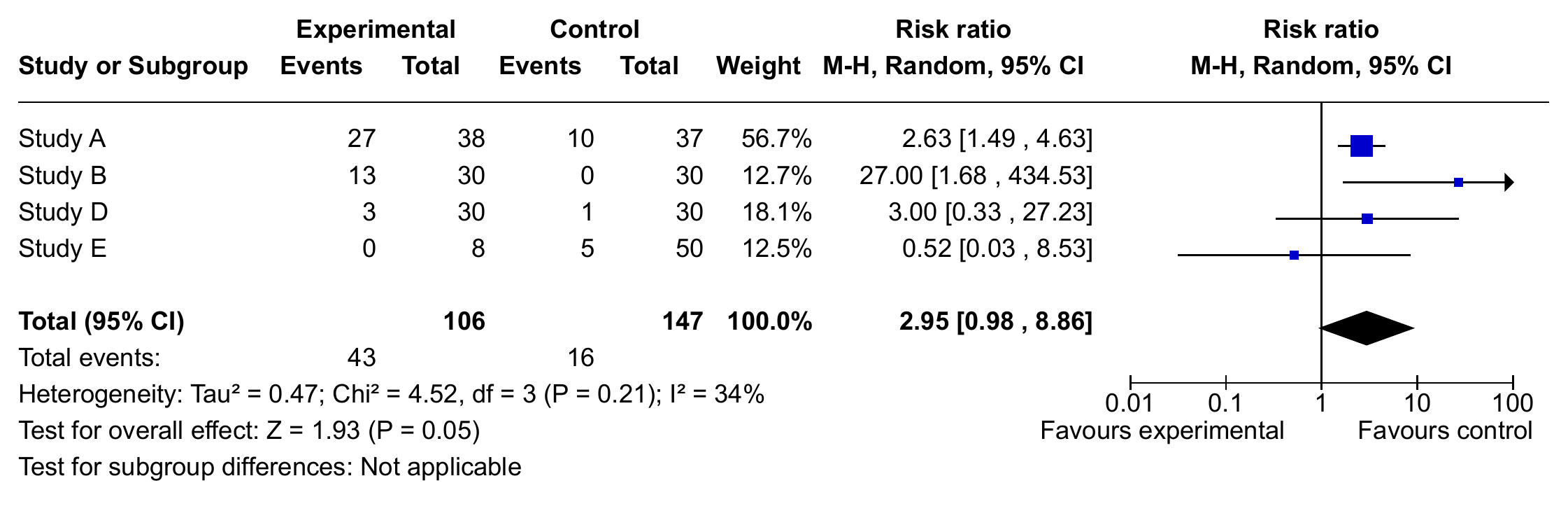}
    \caption{Not including study C will overestimate the review outcome, yet it will be within the 95\% CI range.}
    \label{fig:res_c}
\end{subfigure}
\begin{subfigure}[t]{0.48\textwidth}
    \includegraphics[width=\textwidth]{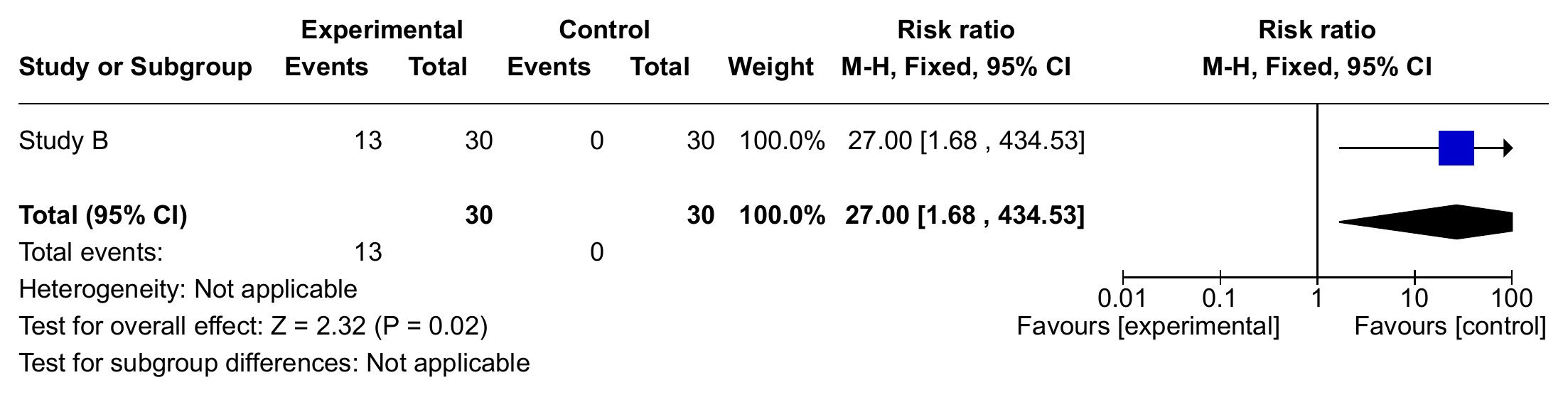}
    \caption{Not including studies A, C, D and E will overestimate the review outcome, and it will be above the 95\% CI range of the original outcome.}
    \label{fig:res_e}
\end{subfigure}
\begin{subfigure}[t]{0.48\textwidth}
    \includegraphics[width=\textwidth]{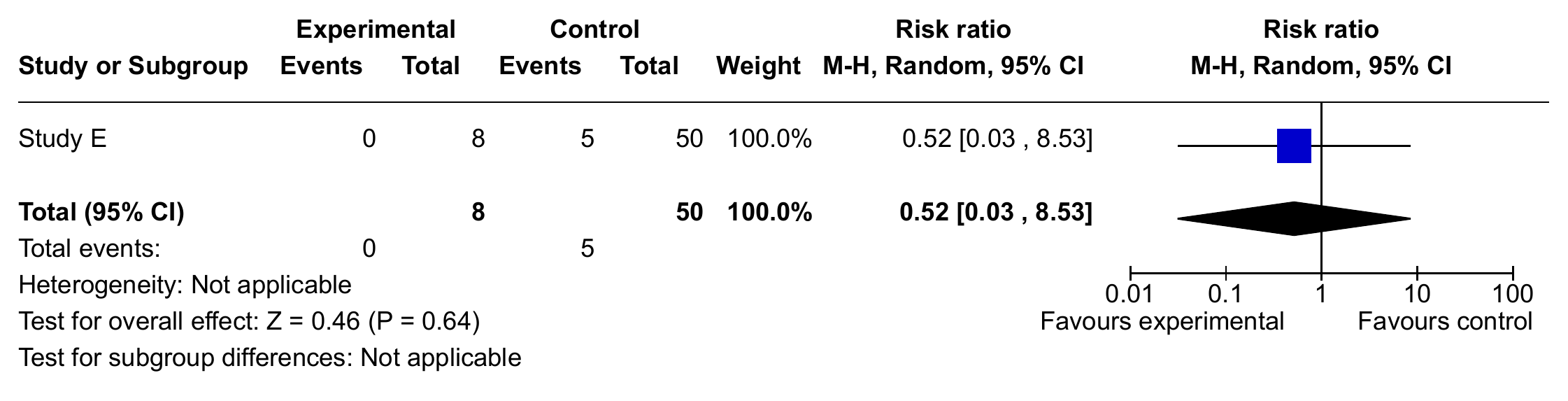}
    \caption{Not including studies A, B, C and D will change the study outcome -- from `favours control' to `favours experimental'.}
    \label{fig:res_f}
\end{subfigure}
\begin{subfigure}[t]{0.48\textwidth}
    \includegraphics[width=\textwidth]{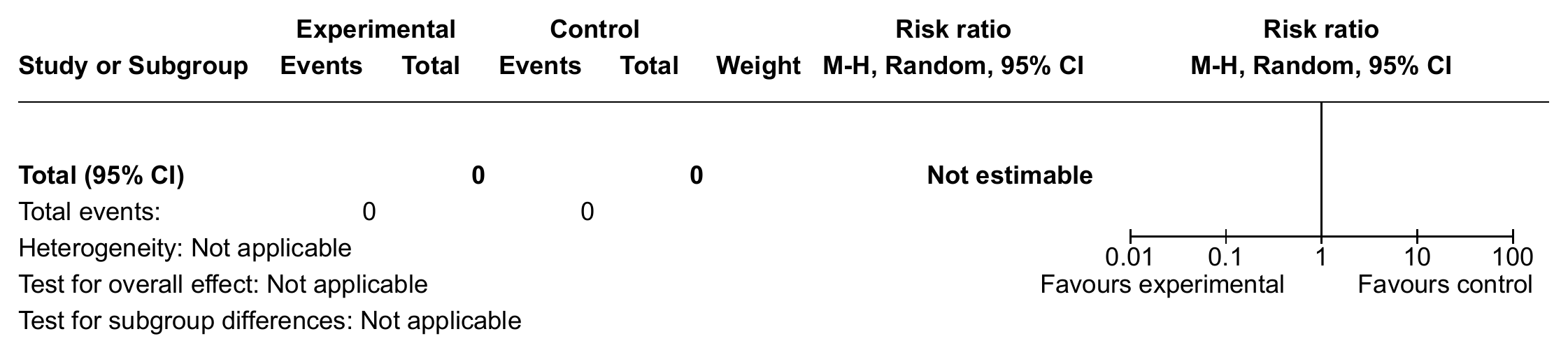}
    \caption{Not including any study makes the outcome non-estimable.}
    \label{fig:res_g}
\end{subfigure}
\caption{Different versions of review outcomes represented as forest plots. Each row is a single study. Columns from the right represent, respectively: (1) the study identifier, (2) number of events in the experimental group (e.g., patients with specific symptoms or adverse events), (3) experimental group size, (4) number of events in the control group, (5) control group size, (6) the weight of a study, and (7) effect size of a study: a difference (e.g., risk ratio or standardised mean difference) in events between experimental or control group. Simulations and figures done using RevMan Web, available at \url{http://revman.cochrane.org}.} %
\label{fig:variants}
\end{figure*}

We analyse the results by examining the outcomes generated by the run and compare them with the outcomes obtained by the original review (Figure~\ref{fig:variants}). 
We extend the analysis done by \citet{NUSSBAUMERSTREIT20181}, who proposed two categories of ``changed conclusions'': (1) if the new review drew the opposite conclusion, (2) if it is not possible to draw a conclusion or the new conclusion has less certainty.
We distinguish five aspects of analysis for review outcomes against the original review (Figure~\ref{fig:res_a}).
The first two of these aspects are real-valued, whereas the remaining three are categorical:

\begin{enumerate}[leftmargin=14pt]
    \item {\emph{Magnitude of difference}} --- By how much are the outcomes different in their effect size (Figure~\ref{fig:res_a} versus~\ref{fig:res_b})? In other words, what is the numerical \emph{impact} on the review outcome when certain studies are not included? This is measured by calculating the relative difference in effect size between the original outcome~$O_{o}$ and predicted outcome~$O_{p}$: $MoD = \frac{\| O_{o} - O_{p} \|}{\|O_{o}\|}$. When $O_{o} = 0$ and $O_{p} \ne 0$, we assume $MoD = 100\%$; otherwise, when $O_{o} = O_{p} = 0$, we set $MoD = 0\%$. Similarly, when the predicted outcome cannot be estimated, we assume $MoD = 100\%$.
    \item {\emph{Distance from CI}} --- Is the new outcome within the Confidence Interval (CI) of the original outcome (Figure~\ref{fig:res_c})? The answer is a distance between the predicted outcome $O_{p}$ and the closest of the pair ($CI_{lower}, CI_{upper}$):%
     $$ \Delta_{CI} = \begin{cases} \| O_{p} - CI_{lower} \| & \text{if $O_{p} < CI_{lower}$,} \\
     \| O_{p} - CI_{upper} \|  & \text{if $O_{p} > CI_{upper}$,} \\
 0 & \text{otherwise.} \end{cases} $$
    \item {\emph{Overestimation/underestimation}} --- Is the outcome overestimated or underestimated compared to the original one (Figure~\ref{fig:res_e})? We first check if the calculated outcome is equal (due to the limits of precision of data reported in RevMan files, we use the relative and absolute tolerance of $10^{-5}$ and $10^{-6}$ respectively). Then, if the outcome is different, we check if the result is higher than the original (overestimation) or lower (underestimation). The answer has three options: ``overestimated'', ``underestimated'', and ``equal''. 
    \item {\emph{Sign}} --- Does the outcome have the same sign as the original one (Figure~\ref{fig:res_f})? In other words, are the new conclusions opposite to the original ones? The answer is binary:  ``yes''/``no''. This aspect corresponds to the first category from \citet{NUSSBAUMERSTREIT20181}.
    \item {\emph{Estimability}} --- Is it possible to calculate the outcome (Figure~\ref{fig:res_g})? An outcome cannot be calculated if there are no included studies concerning it. The answer is binary: ``yes''/``no''.
\end{enumerate}

\section{Experiment Setup}

Contrary to the traditional evaluation based on retrieving relevant publications, with our framework we envision the evaluation in an outcome-based approach.
Specifically, 
we do not treat a dataset as a collection of systematic reviews but rather a collection of outcomes. 
The problem of conducting a systematic review is multi-dimensional.
One can think of it as having several outcomes reporting different dimensions of the review, and the evaluation of the user's needs is conducted independently from each outcome's perspective.
We do not want to average across reviews, each containing a different number of outcomes. We add or average these outcome-level results instead.

Before we present the results, we first discuss the used dataset and models. 

\subsection{Dataset and Models} \label{sec:models}

We used a collection of 38 systematic reviews of interventions from the CLEF TAR 2019 training and test datasets~\cite{Kanoulas2019CLEF2T}. 
Each review consists of a Cochrane ID, a protocol, and a list of publications described by their PubMedIDs with qrels both on the title and abstract level and a full-text level. 
We enhanced the dataset by collecting RevMan files and information about the data and analysis as described in Section \ref{sec:data_extraction}.

Out of 38 reviews in CLEF TAR 2019, our script found studies and outcomes for 32 reviews (17 in the training subset and 15 in the test subset).
We summarise the statistics of the 32 reviews we consider in Table \ref{tab:data_statistics}.
There is a significant discrepancy in the number of outcomes reported by the reviews, ranging from as few as 2 or 3 outcomes in small reviews to 128 outcomes in the largest one. 
Moreover, the majority of these outcomes come from just one or two studies, which presents an additional challenge.

These 32 reviews report 1640 included publications, out of which we managed to find PubMed IDs for 1175 of them (71.6\%).
Next, we wanted to match publications identified with our script to the CLEF TAR 2019 qrels based on the PubMed ID.
There were, in total, 778 relevant documents on the full-text level identified in the CLEF TAR for these 32 reviews.
We successfully merged 741 publications (95.2\% of the total in CLEF TAR); there are only 37 publications in CLEF TAR 2019 qrels which we do not have in our records.

\begin{table}[t]
\centering
\caption{Statistics of the considered dataset. 
}
\label{tab:data_statistics}
\begin{tabular}{@{}llcccc@{}}
\toprule
            &               & \multicolumn{4}{c}{CLEF TAR 2019} \\ 
\multicolumn{2}{l}{Dataset split}                 & \multicolumn{2}{c}{Training}           &    \multicolumn{2}{c}{Test}              \\ \midrule
\multicolumn{2}{l}{Reviews' type} &  \multicolumn{4}{c}{\ \ --- Interventional --- } \\ 
\multicolumn{2}{l}{\# Reviews}                 & \multicolumn{2}{c}{17}       & \multicolumn{2}{c}{15}              \\ \midrule
\multirow{3}*{Outcomes per review} & \hspace{-0.1cm}Min              & 2            &   & 3            &                 \\
& \hspace{-0.1cm}Median           & 9           &   & 15           &                 \\
& \hspace{-0.1cm}Max              & 41          &   & 128          &                 \\ \midrule
\multirow{3}*{Studies per outcome} & \hspace{-0.1cm}Min    & 1            &   & 1            &                 \\
& \hspace{-0.1cm}Median  & 2            &   & 2            &                 \\

& \hspace{-0.1cm}Max    & 55            &   & 40            &                 \\ \bottomrule

\end{tabular}
\end{table}

We use 34 official CLEF TAR 2019 runs from three teams. 
The teams used a variety of ranking methods, including traditional BM25, interactive BM25, continuous active learning, relevance feedback, and various stopping criteria.
Additionally, we included 40 runs based on the reproducibility of the active learning method by~\citet{yang2022goldilocks}.
In total, we evaluate 74 runs, but for the sake of brevity, in this paper, we present the results on a subset of 28 runs, as some of the runs were very similar to each other.
Our model requires full-text assessments, and thus, we use qrels from the full-text level, despite the fact that runs have been trained on titles and abstracts.
While this might not be fair towards the evaluated systems, our experiments aim not to establish which systems are better but to provide an example of the operationalisation of our framework and its implications.

\section{Outcome-based Evaluation} \label{sec:outcome-based-eval}

We first run a simulation study to understand the results of our evaluation framework better in a controlled manner.
Then, we discuss the usage of the evaluation framework with retrieval and classification runs on CLEF TAR 2019 collection.

\subsection{Preliminary Simulation}

We are interested in executing a preliminary study to understand the effect our outcome-oriented evaluation has on the analysis of systematic review automation methods.

\begin{table*}[h]
    \centering
    \caption{Initial results of the simulation on the publication level. Outcomes are aggregated across 32 systematic reviews and are averaged from 20 different random seeds.}
    \label{tab:publication_results}

\resizebox{0.95\textwidth}{!}{
\begin{tabular}{llrrrrrrrrrrrr} 
\toprule
& & &  \multicolumn{11}{c}{N relevant \textbf{publications} removed from the review} \\ 
\multicolumn{2}{l}{Analysis Aspect} & gold & 1 & 2 & 3 & 4 & 5 & 10 & 15 & 20 & 30 & 50 & 100 \\ \midrule 
1 & Mean relative difference & 0.0 & 0.9 & 2.5 & 5.3 & 7.1 & 10.0 & 18.3 & 26.2 & 36.5 & 54.9 & 65.5 & 84.5 \\ \midrule
2 &  Mean distance from CI & 0.000 & 0.002	& 0.003 &	0.004 &	0.007 &	0.008	& 0.013	& 0.042	& 0.102	& 0.018	& 0.008	& 0.083 \\ \midrule
\multirow{4}*{3} &  Equal outcome &  824 & 786 & 750 & 706 & 657 & 623 & 496 & 410 & 340 & 256 & 164 & 80 \\ 
 &  Different & 0 & 38 & 73 & 117 & 167 & 200 & 328 & 413 & 483 & 567 & 659 & 743 \\ 
 &  \hspace{0.3cm}- Underestimated &  0 & 17 & 27 & 38 & 57 & 66 & 98 & 103 & 90 & 55 & 58 & 23 \\ 
 &  \hspace{0.3cm}- Overestimated & 0 & 20 & 45 & 79 & 109 & 134 & 229 & 309 & 393 & 512 & 601 & 720 \\  \midrule
\multirow{2}*{4} &  Have same sign &  824 & 815 & 800 & 774 & 756 & 735 & 663 & 597 & 516 & 365 & 277 & 121 \\ 
 &  Have different sign & 0 & 9 & 24 & 49 & 67 & 88 & 160 & 227 & 307 & 458 & 546 & 702 \\ \midrule
\multirow{2}*{5} & Reported outcomes  & 824 & 816 & 804 & 781 & 767 & 743 & 675 & 610 & 529 & 371 & 284 & 128 \\ 
& Missing outcomes & 0 & 7 & 20 & 43 & 56 & 80 & 148 & 213 & 294 & 452 & 539 & 695 \\ \midrule \midrule
\multicolumn{2}{l}{Average $Recall$ for publications} & 1.00 & 0.92 & 0.84 & 0.75 & 0.70 & 0.63 & 0.45 & 0.35 & 0.28 & 0.22 & 0.14 & 0.05 \\
\multicolumn{2}{l}{Average $Recall$ for studies} & 1.00 & 0.97 & 0.91 & 0.80 & 0.77 & 0.68 & 0.53 & 0.43 & 0.37 & 0.31 & 0.22 & 0.12 \\
\bottomrule
\end{tabular}
}

\end{table*}

We simulate the evaluation framework by taking the set of included \emph{publications} for each review and randomly removing $[1, 2, 3,$ $4, 5, 10,$ $15, 20, 30,$ $50, 100]$ publications from the set and then re-calculating the outcomes. In other words, we are interested in exploring the impact of false negatives on the final review outcome. We compare the outcomes with the `gold' outcomes from the original review. Results from all 32 systematic reviews are reported in Table \ref{tab:publication_results}. 
In our analysis, we consider the metrics from all five analysis aspects (Section~\ref{sec:results-analysis}), as well as the Recall.

\begin{figure}[b!]
\centering
    \includegraphics[width=\columnwidth]{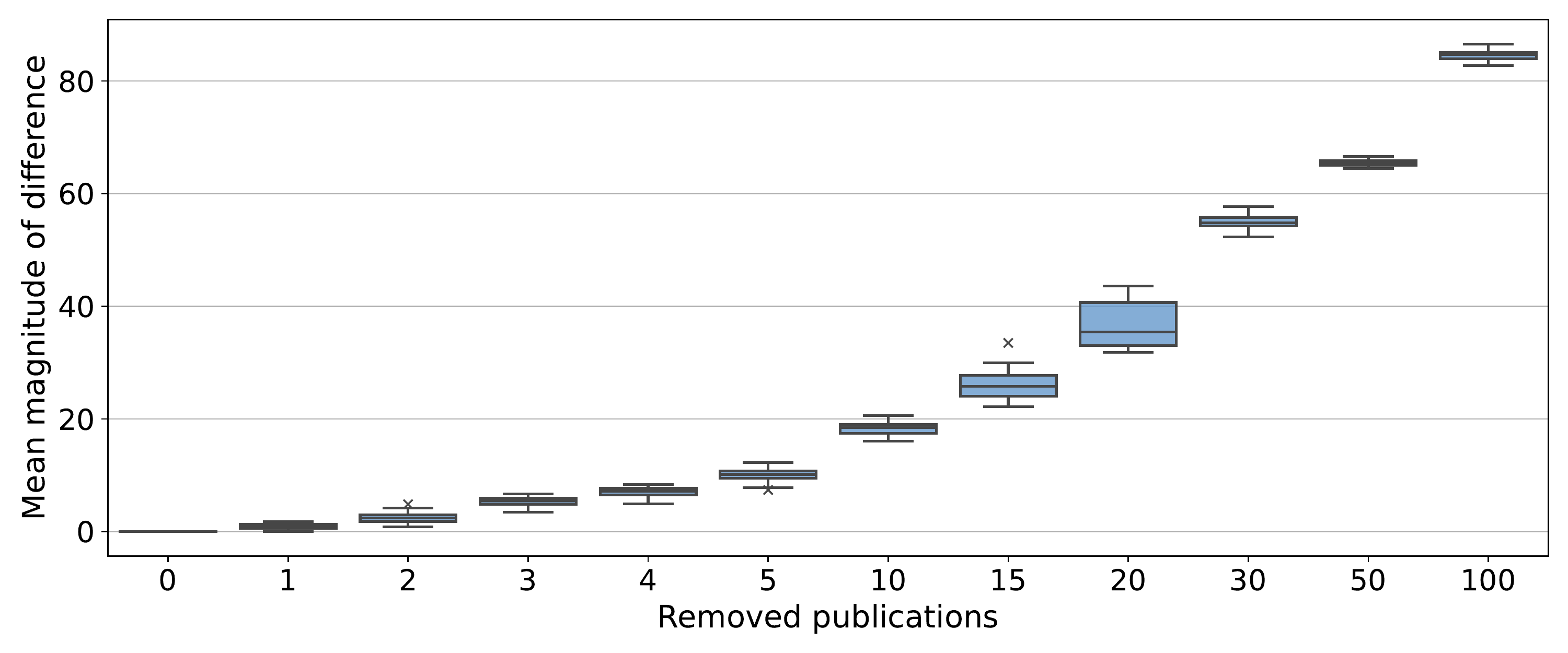}
\caption{Box plots presenting relative difference values from 20 simulations on the publication level. Note that the x-axis does not preserve the linear step.}
\label{fig:simulation_boxplots}
\end{figure}

\begin{figure}[ht]
\centering
    \includegraphics[width=\columnwidth]{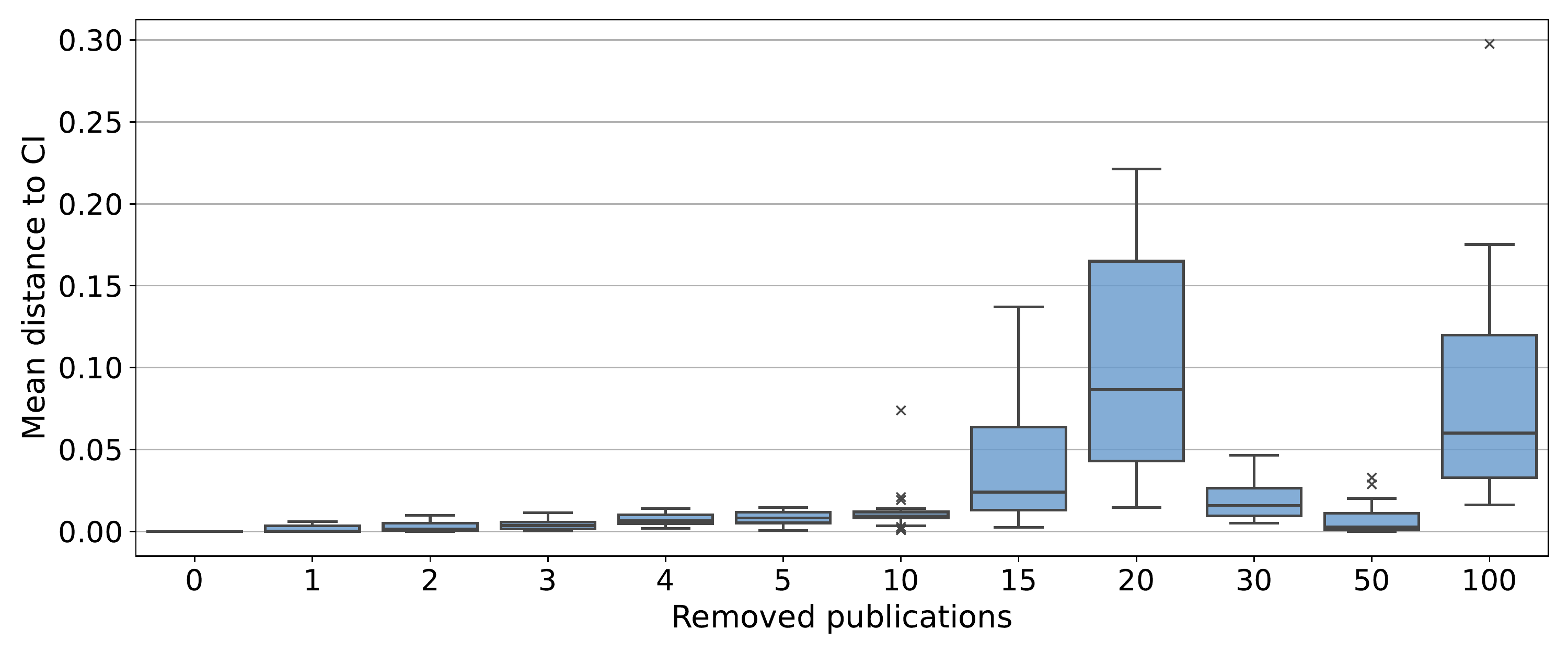}
\caption{Box plots presenting distance to confidence intervals values from 20 simulations on the publication level. Note that the x-axis does not preserve the linear step.}
\label{fig:simulation_boxplots_distance_CI}
\end{figure}

Figure \ref{fig:simulation_boxplots} presents box plots of averaged relative difference (aspect~(1)) values from our simulation at a cut-off at 20\% of the total number of documents.
These results validate our expectations regarding the behaviour of this aspect of analysis as the relative difference grows with the number of removed publications.
On the other hand, the distance to confidence intervals (aspect~(2), Figure~\ref{fig:simulation_boxplots_distance_CI}) does not show any specific trend on the CLEF 2019 reviews.

Out of all the metrics, the one that changes the most when varying the number of removed publications is estimability (5). As more publications are removed, it becomes more and more challenging to calculate outcomes, predominantly because half of the original outcomes relied on one or two studies. 
At the very extreme, when 100 publications are removed from every review, only 15\% of outcomes are still estimable. %

The measure of overestimation and underestimation (3) is showing growing trends with more publications being removed.
Already not including one publication per review (achieving an average recall of 92\% for publications and 97\% for studies) changed 38 outcomes (4.6\% of the total number of outcomes). 
This shows that the commonly used threshold of 95\% Recall does not enforce preserving the same outcomes of the review.
We also notice that the sign (4) aspect is not very descriptive across the simulations as it is mainly influenced by non-estimable outcomes.

\subsection{Evaluation with actual runs}

In this section, we use the prediction on the test subset of the dataset from runs described in Section~\ref{sec:models} and evaluate them using our framework.
We further consider two baselines:

\begin{description}
    \item[gold] -- the best possible run which returns all relevant studies from the original review first. 
    \item[max-with-qrels] -- this run takes into account the limitations of the CLEF TAR collection and our PubMed articles matching process. It uses all relevant studies identified in the CLEF TAR 2019 qrels as relevant and places them first.
\end{description}

We follow the evaluation procedure of CLEF TAR and calculate the following traditional evaluation measures: Mean Average Precision ($MAP$), last relevant found, Recall@k\% of top-ranked publications, with k in [5, 10, 20, 30, 50], Work Saved over Sampling at r\% of recall with r in [95\%, 100\%] ($WSS@95\%$, $WSS@100\%$), $nDCG@20\%$ of top-ranked publications and Area Under Recall Curve ($AURC$).
CLEF TAR as their primary reporting measure used $MAP$; therefore, we will treat $MAP$ as the reference measure when sorting runs.
We do not evaluate baselines with traditional measures, yet for the purpose of sorting, we assume that they achieved the highest MAP score.

We calculate the relative difference in study outcomes (analysis aspect (1) in Section \ref{sec:results-analysis}) for every outcome in all reviews.
The lower the average score is, the better the runs, as their effect differs less from the original review effect.
As considered runs were rankings, we follow the same procedure as for Recall and nDCG, namely we calculate the relative difference at k\% of top-ranked publications with k in [5, 10, 20, 30, 50].

\begin{figure}[h]
\centering
    \includegraphics[width=1\columnwidth]{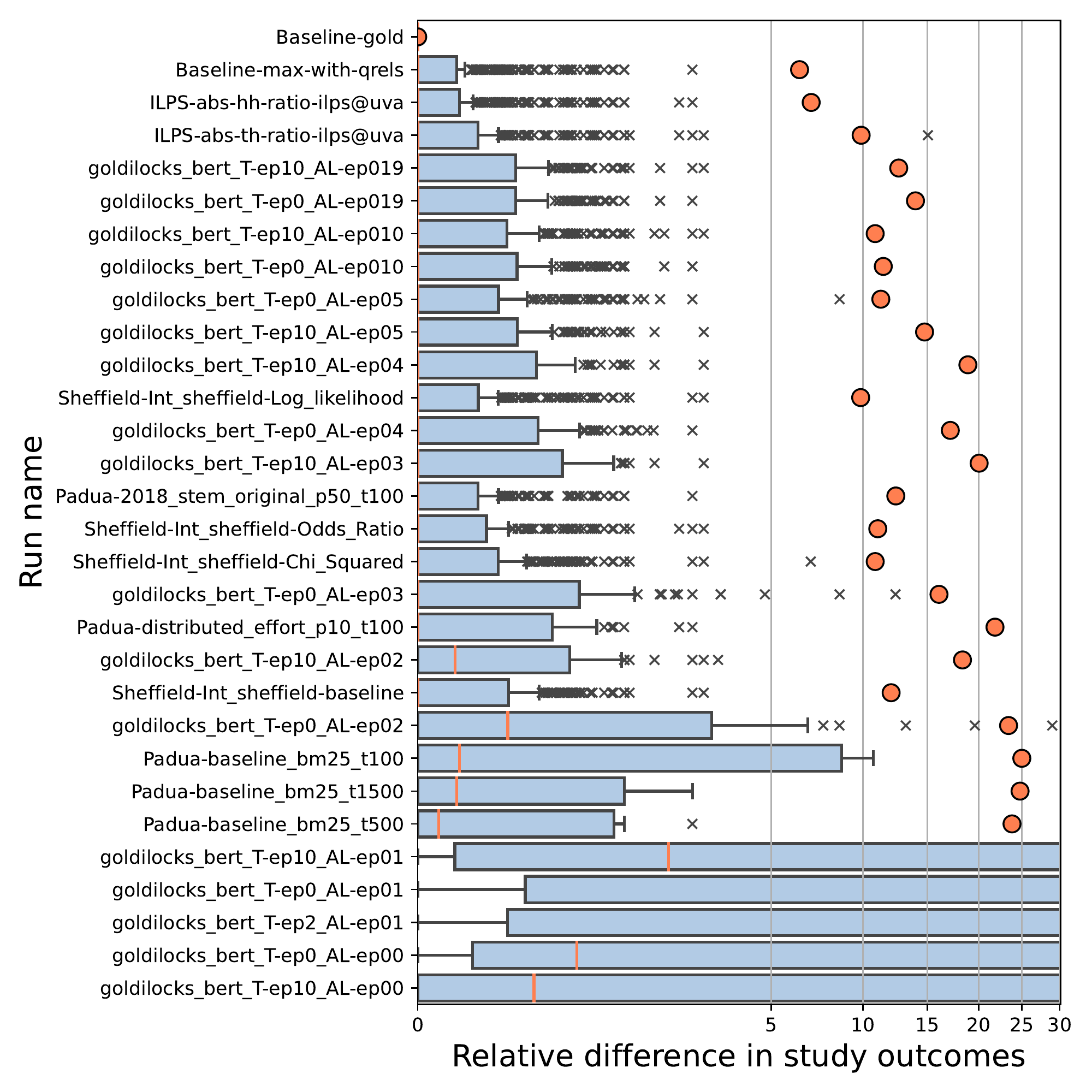}
\caption{Box plot presenting runs with their relative difference in study outcomes for an evaluation with a cut-off at 30\% of the total number of documents for each review. Runs are sorted by their MAP score. The orange circle denotes the mean relative difference $\mathbf{@30\%}$. The X-axis is cut at 30, while the outliers exist up to the value of 100; we cut for visualisation purposes.}
\label{fig:mod_per_run}
\end{figure}

\begin{figure}[ht]
\centering
    \includegraphics[width=\columnwidth]{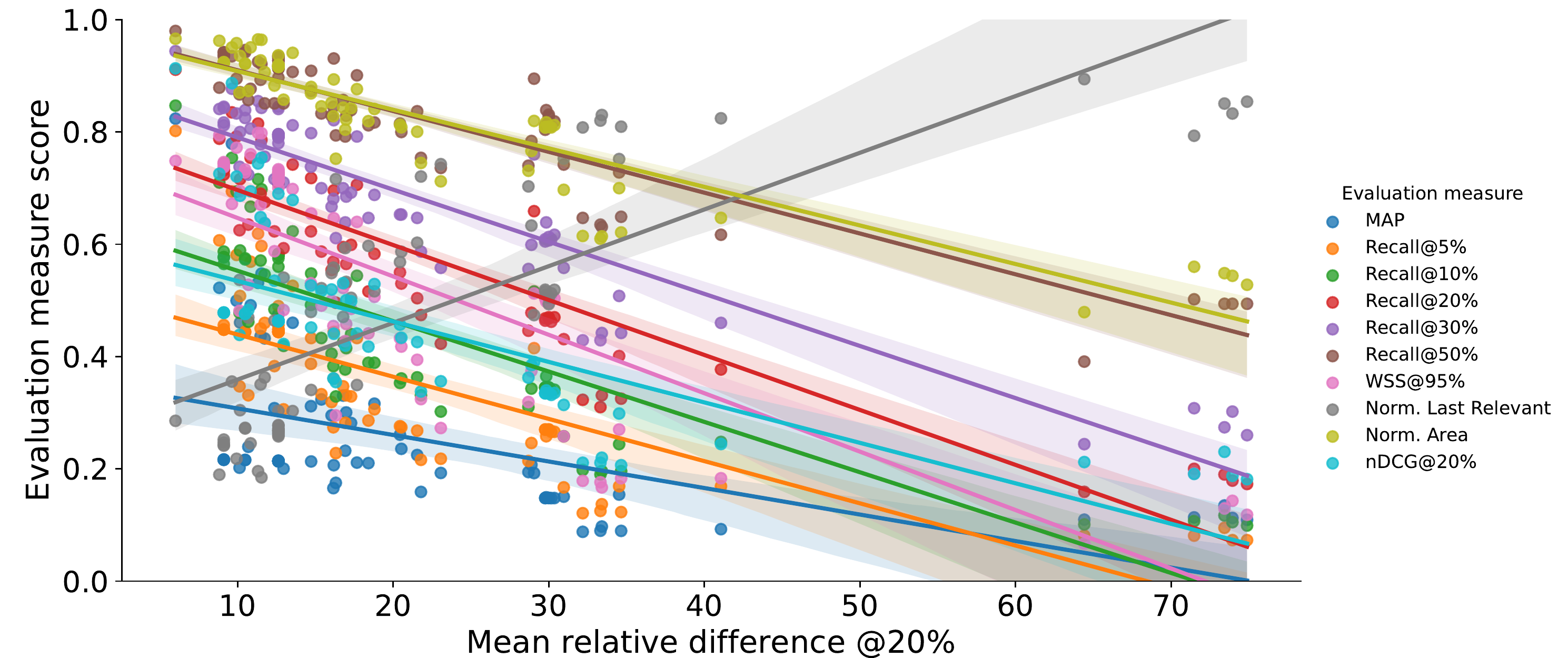}
    \caption{Linear regression fits between relative difference at 20\% cut-off of documents and other evaluation measures scores. Correlations for relative difference at other cut-offs follow similar trends.} 
\label{fig:evaluation_correlation}
\end{figure}

Figure~\ref{fig:mod_per_run} presents a box plot of relative difference per outcome calculated at 30\% cut-off of dataset size for 15 test CLEF TAR reviews.
Except for the best run, all other runs changed their rank when ordered using their mean relative difference score compared to the MAP-based ranking.
While top runs, according to MAP scores, have low variability, there are runs among the top 10 which show considerable fluctuation.
This means there are specific reviews for which these runs will lead to significantly different decisions about the outcome.
This behaviour is comparable for relative difference at other cut-offs $k$.

What is also interesting is that the mean relative difference at~$30\%$ cut-off for the \emph{max-with-qrels} baseline run equals 6.24.
Furthermore, for the relative difference score calculated at 100\% of documents, this baseline score is also not equal to 0.
This means that the limitations of the CLEF TAR collection and qrels establish a lower bound for the best achievable value of relative difference.

Figure~\ref{fig:evaluation_correlation} presents correlation between relative difference calculated at 20\% cut-off 
 of dataset size and evaluation measures used at CLEF TAR 2019.
The score correlates positively with the last relevant found, but there is a negative correlation with all other measures.
This confirms our intuition that a higher average relative difference score across outcomes means a worse model effectiveness, as the ideal `best' model should achieve a difference of 0.

\subsection{Pareto Frontier Optimisation}

Based on the simulation results, we note a problem with non-estimable outcomes. 
Should these outcomes be assigned a zero score or maybe an infinite value? 
This raises the issue of handling these values in the evaluation process for calculating relative difference scores.
In our study, we assigned a zero value to non-estimable outcomes, which allowed us to assume that the relative difference equals 100\%.
Nevertheless, this yields the problem of when the actual outcome is equal to the zero value (i.e., the study does not favour the experimental nor the control group), as the difference, in this case, would also be zero.
One way to overcome the issue of non-estimable outcomes would be to evaluate both estimability and relative difference implemented, for instance, using the Pareto frontier~\cite{lotov2008visualizing}.

\begin{figure}[t]
    \centering
    \includegraphics[width=\columnwidth]{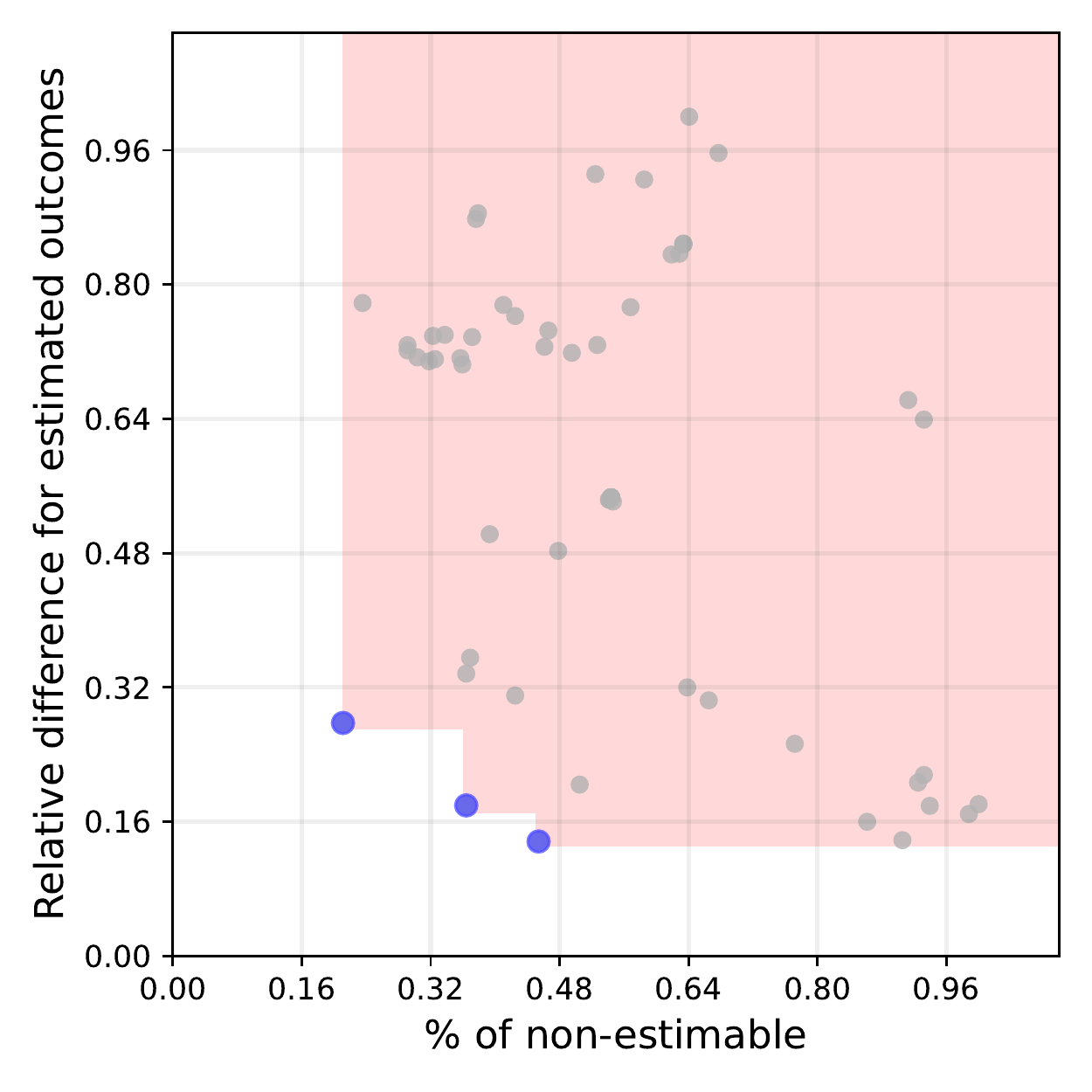}
    \caption{Visualisation of the Pareto frontier for two objectives: (1)~number of non-estimable outcomes on the x-axis and (2)~sum of relative difference for estimable outcomes on the y-axis. Both objectives are to be minimised. Runs are evaluated at a cut-off at 5\% of the total number of documents for each review. Non-dominated runs are marked with a blue colour.}
    \label{fig:pareto_frontier}
\end{figure}

Figure \ref{fig:pareto_frontier} presents the Pareto frontier %
evaluated at a cut-off at 5\% of the total number of documents.
On the x-axis, we show the number of non-estimable outcomes for each run.
On the y-axis, there is a sum of relative difference for estimable outcomes.
We min-max normalise the sums including the gold baseline run (gold represents the best achievable score of $(0,0)$).
Both objectives should be minimised, i.e., we want to have as few non-estimable outcomes as possible and for all estimated outcomes, the difference would be as close to zero as possible.
Contrary to the previous evaluations, we can notice that no single run would dominate on both dimensions. 

\section{Limitations}

The primary objective of this paper was to introduce the concept of evaluating automated methods for systematic reviews based on their impact on review outcomes, rather than relying on binary qrels.
In this section, we reflect on the potential limitations that arise when attempting to fully operationalise our proposed framework.

\textbf{Do not optimise models using this measure.}
A practice that can be observed across the field is treating evaluation measures as an optimisation objective.
We believe that our evaluation approach should not be used for optimising models.
The notion of difference in study outcomes is only known a-posteriori when the review is completed.
Using absolute differences in study outcomes as an optimisation objective might lead to over-fitting to biases in data.

\textbf{Other types of systematic reviews.} 
We focus only on systematic reviews of interventions which have a clear structure and evaluate the effectiveness of specific treatments, programs, or policies by comparing experimental setups with control groups.
However, there are several other types of systematic reviews, such as diagnostic test accuracy reviews, prognostic reviews, and qualitative research reviews, each of which presents unique challenges for automation and evaluation~\cite{Kanoulas2019CLEF2T}. Future work should investigate how this outcome-based evaluation framework can be extended to these other types of reviews.

\textbf{Different outcome types.} 
While our proposed evaluation framework focuses on continuous and dichotomous outcomes, other types of outcomes may be reported in systematic reviews, including ordinal, count, and time-to-event data. In our analysis, however, we found that continuous and dichotomous outcomes comprised most of the outcomes in the dataset we studied, accounting for 92\% of all reported outcomes across 32 CLEF TAR 2019 reviews.

We believe that our evaluation framework could be generalised to incorporate other types of outcomes.
Additionally, while we attempted to closely follow the evaluation protocols from the Cochrane handbook, some shortcuts were taken during the implementation process (for 2.4\% of outcomes our effect calculations yielded marginally different results). %
In future work, ideally, access to RevMan or another official program for calculating study outcomes would be needed to make sure that all outcome types are covered.

\textbf{Title and abstract screening.} 
We work on the outcomes extracted from the full-text screening and use relevance judgments from full-text screening to judge the runs.
However, most models are trained on titles and abstracts, which might make this an unfair comparison.

\section{Conclusion}

This paper puts forward a novel, outcome-based evaluation framework for assessing the effectiveness of automatic search strategies and citation screening methods in the context of systematic literature reviews. 
Our proposed framework evaluates the quality of these methods based on how closely the outcomes of their included publications match the actual review outcomes. 
We believe that this approach offers a more accurate reflection of real-world scenarios where not all included publications have the same impact on the final review outcome.

In addition to proposing the framework, we explore five analysis aspects that it enables, including measuring the numerical difference in predicted systematic review outcomes.
We run initial experiments to simulate the impact of false negatives on reviews' outcomes showing that five missing publications per review can change 24\% of outcomes. 
We also compare the evaluation results obtained using our framework with those obtained using traditional evaluation methods on CLEF TAR 2019 runs, highlighting the differences in focus between the two approaches.

Overall, we believe this framework represents a step forward in developing more effective and realistic methods for evaluating automation methods in the context of systematic literature reviews in medicine and in other domains in which the importance of systematic reviews is increasing.

\begin{acks}
This work was supported by the EU Horizon 2020 ITN/ETN on Domain Specific Systems for Information Extraction and Retrieval -- DoSSIER (H2020-EU.1.3.1., ID: 860721).
This research is partially funded by the Australian Research Council, Discovery Project DP210104043. 
\end{acks}

\bibliographystyle{ACM-Reference-Format}
\balance
\bibliography{references}

\end{document}